\documentclass{article}
\usepackage[english]{babel}
\newcommand{\bb}{\begin{equation}}
\newcommand{\ee}{\end{equation}}

\begin{document}

\thispagestyle{empty}

\begin{center}
{\Large\bf
Conformally-flat Stackel spaces in Brans-Dicke theory}

\bigskip
{\bf V.V. Obukhov\footnote{obukhov@tspu.edu.ru} , K.E. Osetrin\footnote{osetrin@tspu.edu.ru},
  A.E. Filippov\footnote{altair@tspu.edu.ru}  , Yu.A. Rybalov\footnote{ribalovyua@tspu.edu.ru}}

\bigskip
{\it Tomsk State Pedagogical University, Tomsk, Russia}
\end{center}

\bigskip
\begin{abstract}
The classification problem for conformally-flat space-times that admit a separation of variables in the Hamilton-Jacobi equation of the scalar-tensor Brans-Dicke theory of gravity is examined. The field equations of the scalar-tensor theory of
Brans and Dicke for conformally-flat Stackel space-times of type (1.1) are solved. An explicit form of the metric tensor and scalar field is obtained.
\end{abstract}

\bigskip
\section[Introduction.]{Introduction.}

Conformally-flat spaces are widely used in cosmology as simple but sufficiently realistic models of space-time. In particular, they include the models of de Sitter and Freedman-Robertson-Walker. On the other hand, the available experimental data evidence that the theory of general relativity (GRT) is insufficient for a construction of adequate models. This fact stimulates a search for alternative theories of gravity.
The gravity theory of Brans-Dicke (BDT)  \cite{Hunta1,Hunta2} is one of the first scalar-tensor theories of gravity. At present, scalar-tensor theories are of interest as low-energy approximations of quantum field theories. BDT differs from GRT by the form of its field equations; however, the influence of gravity on physical systems in BDT and GRT is equivalent, and is determined by the metric tensor. In both BDT and GRT, test particles move along geodesic lines, and therefore in BDT one is interested in Stackel spaces (SS), which, by definition, admit integration of the equations of motion for test particles in the form of Hamilton-Jacobi, according to the method of a complete separation of variables. In finding exact solutions of the field equations in BDT, Stackel spaces, however, are more important than they are in GRT, because in these spaces there is a possibility of finding a general solution of the scalar equation that enters the set of field equations of BDT.
In this article, we examine conformally-flat spaces that admit a complete separation of variables in the Hamilton-Jacobi equation of the Brans-Dicke theory. Conformally-flat metrices of this kind, the Stackel metrices of type (1.1), (2.1) and (3.1), have been found in  \cite{Hunta3}.

\section{General information on conformal SS of type (1.1).}

Let us recall that Stackel spaces are such Riemann spaces that admit an integration of the equations of geodesic lines by the method of a complete separation of variables in the Hamilton-Jacobi equation. A more detailed description of Stackel spaces and their classification can be found, e.g., in \cite{Hunta4}.

We recall that conformally-flat spaces are such spaces whose metrices satisfy the equation $C^i_{jkl}=0$, where $C^i_{jkl}$  is the Weyl tensor.

Let us present the metrices of the conformal SS of type (1.1) obtained in \cite{Hunta3}, namely,
\begin{equation}
ds^2=\Delta^2(2W^{(1)}(x^2,x^3)dx_0dx_1+G(x^2,x^3)d{x^1}^2-\epsilon_2 W^{(2)}(x^0,x^3)d{x^2}^2-\epsilon_3 W^{(3)}(x^0,x^2)d{x^3}^2),
\end{equation}
where
$$W^{(2)}(x^0, x^3)=t_0 (x^0)+t_3 (x^3),\; W^{(3)}(x^0, x^2)=t_2(x^2)-t_0(x^0),$$ $$ G(x^2,x^3)=g_2(x^2) W^{(2)}+g_3(x^3) W^{(3)},\; \epsilon=\pm 1,
$$
$$W^{(1)}(x^2,x^3)=t_2(x^2)+t_3(x^3)=W^{(2)}(x^0,x^3)+W^{(3)}(x^0,x^2),\; \Delta =\Delta (x^0,x^1,x^2,x^3).$$
The functions $t_0(x^0)$, $t_2(x^2)$, $t_3(x^3)$, $g_2(x^2)$, $g_3(x^3)$ are arbitrary functions of their arguments; $\Delta$ is an arbitrary function of all of its arguments. Here and elsewhere, we assume that constants are denoted by lowercase Greek characters with a subscript or without it; functions of one argument are denoted by lowercase Latin characters with a subscript that denotes the variable; functions of several variables are denoted by uppercase Latin and Greek characters. An exceptional case is given by field components, denoted by uppercase Greek characters with an index that stands for a variable.

The problem of selecting the conformally-flat metrices from the class of conformal SS leads to the condition
$$t_0t_2'=t_0t_3'=0,$$
which implies the presence of two classes of conformally flat
Stackel metrices:
\[\mbox{{\bf Class A.}}\ \  t_0=0,\ \ \ \ \ \    {\bf Class B.}\ \  t_2'=t_3'=0,\]
where the metrics can be presented in the same form for both classes:
\begin{equation} ds^2=\Delta^2(2dx^0dx^1+G(x^2,x^3)d{x^1}^2+Td{x^2}^2+(1-T)d{x^3}^2),\end{equation}
with the functions  and  being determined separately for each class.

{\bf Class A.}
\begin{equation} G(x^2,x^3)=\rho\left( \frac{1}{t_2}-\frac{1}{t_3}\right),\; T=\frac{t_3}{t_2+t_3}, \end{equation}
$$t''_2=\frac{1}{4}t_2{}^2(\mu t_2{}^2+\nu t_2+\kappa),\; \; t''_3=\frac{1}{4}t_3{}^2(-\mu t_3{}^2+\nu t_3-\kappa).$$

{\bf Class B.}
\begin{equation} G=-\Big( \frac{\mu {x^2}^2}{2}+\mu_1 x^2 +\mu_2 \Big) t_0-\Big( \frac{\nu {x^3}^2}{2}+\nu_1 x^3\Big)(1-t_0) \Big), \end{equation}
$$T=t_0,\;\; t'_0{}^2=t_0{}^2(1-t_0)^2(2\mu t_0+2\nu(1-t_0)-\kappa t_0(1-t_0)).$$
The condition $g=det\; g^{ij}<0$, imposed on the metrices (2), implies
$$0<T<1,$$
or, equivalently, for the functions $t_2$, $t_3$,
$$t_2t_3>0.$$

\section{Field equations of Brans-Dicke theory.}

The field equations of the Brans-Dicke theory in the case of vacuum have the form \cite{Hunta1}
$$\left\{%
\begin{array}{ll}
    \phi^{\ ;\alpha}_{;\alpha}=0 & \hbox{} \\
    R_{\mu\nu}-\frac{1}{2}g_{\mu\nu}R=-\frac{\omega}{\phi^2}(g_{\mu\nu}\phi^{;\alpha}\phi_{;\alpha}-\phi_{;\mu}\phi_{;\nu})-
    \frac{1}{\phi}(g_{\mu\nu}\phi^{\ ;\alpha}_{;\alpha}-\phi_{;\mu\nu}). & \hbox{} \\
\end{array}%
\right.
$$
Here, $\phi$ is a scalar field; $\omega$ is a constant. A separation of variables in the scalar equation implies the following representation of the function $\phi$:
\begin{equation} \phi=\prod^{n}_{i=1}\phi_i(x^i) \end{equation}
(a multiplicative separation of variables). The form of equations (5) can be simplified, if one excludes the scalar curvature and introduces, instead of $\phi$, a new variable,
\begin{equation} \Phi=ln|\phi|=\sum^{n}_{i=1}\Phi_i(x^i), \end{equation}
which admits a separation of variables of additive type.

Then, equations (5) acquire the form
$$\left\{%
\begin{array}{ll}
    \Phi^{\ ;\alpha}_{;\alpha}+\Phi^{;\alpha}\Phi_{;\alpha}=0 & \hbox{} \\
    R_{\mu\nu}=(\omega+1)\Phi_{;\mu}\Phi_{;\nu}+\Phi_{;\mu\nu} & \hbox{} \\
\end{array}%
\right.
$$
We have the following equations $R_{11},\; R_{12},\; R_{13}$ for the metrics (2):
$$\frac{2\Delta_{,11}}{\Delta}=-\frac{2\Delta_{,1}\Phi_1'}{\Delta}-\omega \Phi_1'{}^2 -\Phi_1'',$$
$$\frac{2\Delta_{,1a}}{\Delta}=-\frac{2\Delta_{,a}\Phi_1'}{\Delta}-\frac{2\Delta_{,1}\Phi_a'}{\Delta}-\omega \Phi_1'\Phi_a',\; \; a=2,3.$$
If one subjects these equations to the following transformation of the unknown function $\Delta$,
$$\Delta \rightarrow \Delta e^{\Phi/2},$$
then the equations take a simplified form:
\begin{equation} \Delta_{,1p}=\omega\Delta\Phi_1'\Phi_p',\; p=1,2,3. \end{equation}

The mentioned change literally coincides with a conformal transformation examined in \cite{Hunta5}; however, in the case under consideration we deal with an arbitrary function , and therefore, the presented change only signifies an expedient choice of the form of equations.
Integrating the subset of equation (9) allows one to obtain a solution of the entire set of equations.

\section{Summary of results}

While searching for solutions, one encounters subclasses to which one applies an indication of vanishing, or non-vanishing, of various components of the scalar field in the additive presentation (7); thus the results are listed in the following classification table:

{\bf Class À:}
\\

\begin{tabular}{|c|c|c|}
\hline
 Form of scalar field & $\Phi_1=0$ & $\Phi_1\neq 0$\\
\hline
 $\Phi_2=\Phi_3=0 $& Subclass A3 & Subclass A2\\
\hline
 $\Phi_2\neq 0,\Phi_3=0 $& - & -\\
\hline
 $\Phi_2\neq 0,\Phi_3\neq 0 $& - &Subclass A1\\
\hline
\end{tabular}
\\
\\

{\bf Class B:}
\\

\begin{tabular}{|c|c|c|}
\hline
 Form of scalar field & $\Phi_1=0$ & $\Phi_1\neq 0$\\
\hline
 $\Phi_2=\Phi_3=0 $& Subclass B5 & Subclass B4\\
\hline
 $\Phi_2\neq 0,\Phi_3=0 $& Subclass B6 & -\\
\hline
 $\Phi_2\neq 0,\Phi_3\neq 0 $& - &Subclass B1, B2, B3\\
\hline
\end{tabular}
\\
\\

\noindent
All the obtained metrices can be presented in the form
$$dS^2=\Omega(2dx^0dx^1+G(x^2,x^3){dx^1}^2+T{dx^2}^2+(1-T){dx^3}^2).$$
In what follows, we also use the notation
$$cos_{\omega}x=\left\{%
\begin{array}{ll}
    cos x,\; if\; \omega<0, & \hbox{} \\
    ch x,\; if\; \omega>0.  & \hbox{} \\
\end{array}%
\right.
$$

\newpage
\noindent
{\bf Class A.}

In all of the solutions for the metrices of class A, there hold relations (3).

{\bf Solution A1.}
\[ \Omega=e^{-\beta\gamma \rho x^0}{cos_{\omega}}^2(\lambda \Phi)e^\Phi,\qquad
 \Phi=\alpha x^0+\beta x^1+\frac 1{\gamma t_2}-\frac 1{\gamma t_3},\qquad \omega=\pm\lambda^2.\]
\[ \nu=-16\,\alpha\beta\gamma^2,\qquad \kappa=-6\beta^2\gamma^2 \rho.\]

{\bf Solution A2.}

\[ \Omega=f_1{}^2\,e^{\Phi_1},\qquad\frac{\,f_1{}''}{f_1}=\omega{\Phi_1'}^2,\]

$\Phi=\Phi_1(x^1)$ is an arbitrary function, $\omega$ is an arbitrary constant.

{\bf Solution A3.}

\[ \Omega=f_0{}^2\,e^{\Phi_0},\qquad\frac {\kappa\rho}{24}+\frac{\,f_0{}''}{f_0}=\omega{\Phi_0'}^2,\]

$\Phi=\Phi_0(x^0)$ is an arbitrary function, $\omega$ is an arbitrary constant.
\\

\noindent
{\bf Class B.}

Everywhere in class B, unless otherwise is specified, there hold relations (4).

{\bf Solution B1.}

\[ \Omega=e^{\Phi}\,{cos_{\omega}}^2(\lambda \Phi),\qquad T=\mbox{const},\qquad\Phi=\alpha x^0+\beta x^1+\gamma x^2+\delta x^3,\qquad
  \omega=\pm \lambda^2.\]
\[ \alpha=\frac 12(\gamma^2(T-1)-\delta^2T).\]

{\bf Solution B2.}

\[ \Omega=\frac{e^{\Phi}\,{cos_{\omega}}^2(\lambda \Phi)}{1-t_0},\qquad
 G=(\alpha-4\,\frac{\gamma^2}{\beta^2}\,{x^2}^2)\,t_0,\qquad t_0{}'=\frac4\beta\,\gamma\,t_0{}^2(1-t_0),\]
\[ \Phi=\Phi_0+\beta\,x^1+\gamma\,{x^2}^2+\delta\,x^3,\qquad
 \Phi_0'=-\frac{\delta^2}{2\beta}+\frac{\delta^2-\alpha\,\beta^2}{2\beta}\,t_0,\qquad   \omega=\pm \lambda^2.\]

{\bf Solution B3.}

\[  \Omega=f_0^2\,{cos_{\omega}}^2(\lambda \Phi)\,e^{\Phi},\qquad
 \frac { f_0{}'}{f_0}=-\frac 12\frac{t_0'}{t_0}+\frac{2\,\gamma}\beta\,t_0,\qquad
 G=(\alpha-4\,\frac{\gamma{}^2}{\beta^2}\,{x^2}^2)\,t_0+
 4\,\frac{\delta{}^2}{\beta^2}\,{x^3}^2(t_0-1),\]
\[ t_0'=\frac 4\beta\,t_0(1-t_0)(\gamma t_0+\delta(t_0-1)),\]
\[ \Phi=\Phi_0+\beta\,x^1+\gamma\,{x^2}^2+\delta\,{x^3}^2,\qquad
 \Phi_0'=-\frac{\alpha\,\beta}2\,t_0 ,\qquad\omega=\pm \lambda^2. \]

{\bf Solution B4.}

\[ \Omega=f_1{}^2\,e^{\Phi_1}\qquad T=\mbox{const},\qquad G=0,\qquad
 \frac{f_1{}''}{f_1}=\omega\,\Phi_1'{}^2, \]

  $\Phi=\Phi_1(x^1),$ is an arbitrary function, $\omega$  is an arbitrary constant.

{\bf Solution B5.}

\[ \Omega=f_0{}^2\,e^{\Phi_0},\qquad
 \frac {f_0{}''}{f_0}=\frac 14\,t_0{}^2\,\biggl(3(2\mu-2\nu-\kappa)+2(-2\mu+2\nu+3\kappa)t_0-
 3\kappa\, t_0{}^2\biggl)+\frac{\nu}2+\omega\,\Phi_0'{}^2,\]

$\Phi=\Phi_0(x^0)$  is an arbitrary function, $\omega$ is an arbitrary constant.

{\bf Solution B6.}

\[ \Phi=\ln \frac{x^2}{\sqrt[3]{t_0f_0{}^2}},\qquad
 \Omega=\frac{f_0{}^2}{x^2}\,e^{\Phi},\]
\[f_0''=-\frac{f_0}{12}(-8\nu+(\kappa+6\nu-2\mu)t_0+(6\kappa+12\nu-17\mu)t_0{}^2+
 5(-3\kappa-2\nu+2\mu)t_0{}^3+  8\kappa\, t_0{}^4)+\frac{f_0'}3(\ln f_0t_0)' ,\]

 $\omega$ is an arbitrary constant.

\section{Conclusion.}

In the present article, we have obtained conformally-flat Stackel spaces in the scalar-tensor gravity theory of Brans-Dicke which admit a complete separation of variables in the Hamilton-Jacobi equations according to type (1.1). A complete form of the metric tensor and scalar field is presented. This work was supported by RFBR Grant N 06-01-00609-a and by President Grant SS-2553.2008.2.

\end{document}